\documentclass[aps,pre,preprint,groupedaddress]{revtex4}

\usepackage{graphicx}
\usepackage{latexsym}
\usepackage{amsmath,amssymb}

\begin{document}


\title{
Bound-state energy
of the $d=3$ Ising model
in the broken-symmetry phase:
Suppressed finite-size corrections
}

\author{Yoshihiro Nishiyama}
\affiliation{Department of Physics, Faculty of Science,
Okayama University, Okayama 700-8530, Japan}

\date{\today}

\begin{abstract}
The low-lying spectrum
of the three-dimensional Ising model
is investigated numerically;
we made use of an equivalence between the
excitation gap and the reciprocal correlation length.
In the broken-symmetry phase,
the magnetic excitations are attractive,
forming a bound state with
an excitation gap $m_2(<2m_1)$ ($m_1$: elementary excitation gap).
It is expected that the
ratio $m_2/m_1$ is a universal constant
in the vicinity of the critical point.
In order to estimate $m_2/m_1$,
we perform the numerical diagonalization
for finite clusters with $N \le 15$ spins.
In order to reduce the finite-size errors,
we incorporated the extended
(next-nearest-neighbor and four-spin) interactions.
As a result, 
we estimate the mass-gap ratio as
$m_2/m_1=1.84(3)$.
\end{abstract}


\maketitle

\section{\label{section1}Introduction}

The first-excitation gap (reciprocal correlation length) is of fundamental
significance
in the theory of critical phenomena.
On one hand,
little attention has been paid to the low-lying spectrum.
Recent consideration on 
the Ising model, however,
revealed rich characters in the low-lying spectrum
\cite{Zamolodchikov98,Fonseca03,Delfino04,Caselle99,Agostini97,%
Provero98,Caselle00,Fiore03,Lee01,Caselle02}.
Actually,
in the broken-symmetry phase,
the magnetic excitations are attractive,
forming a pair of 
bound states with characteristic excitation gap
$\{ m_i \}$.

For example, in two dimensions ($d=2$),
there appear eight types of excitations ($i=1,2,\dots,8$)
\cite{Zamolodchikov98,Fonseca03}.
Each of these excitations possesses an intrinsic mass gap $m_i$.
That is, for each excitation,
the ratio $m_i/m_1$ ($m_1$: an elementary excitation gap)
is a universal constant in the vicinity of the critical point.
[Notably enough,
the elementary excitation $m_1$ is also a composite particle
(bound state) \cite{Delfino04}.
In this sense, all particles are equally
elementary, reflecting a
highly non-perturbative nature of this problem.]
The point is that 
in the continuum (scaling) limit,
the $d=2$ Ising model
is exactly solvable 
even in the presence of the magnetic field;
see Ref. \cite{Delfino04} for a review.
As mentioned afterward,
(properly scaled) 
infinitesimal magnetic field is significant to 
stabilize the bound state.

On the contrary, in $d=3$,
such an exact solution is not available.
The detailed structure of the $d=3$ spectrum is not fully understood.
So far,
the lowest bound state $m_2$, namely, the second-excitation gap,
has been studied in detail 
\cite{Caselle99,Agostini97,Provero98,Caselle00,Fiore03,Lee01,Caselle02};
a
series of bound states is considered in 
Refs. \cite{Caselle99,Caselle00,Fiore03,Caselle02,Fisher71}.
For example,
by means of the Monte Carlo method \cite{Caselle99},
a clear indication of a bound state
$m_2$ was found.
The simulation data indicate the mass-gap ratio
\begin{equation}
m_2/m_1=1.83(3)     .
\end{equation}
(See also Refs. \cite{Agostini97,Provero98,Caselle00,Fiore03}.)
Note that the Monte Carlo method
does not yield the excitation gap $m_2$ directly.
Rather,
one has to look into the asymptotic form of the correlation function,
and estimate the sub-dominant decay rate $\propto 1/m_2$
together with the dominant one $\propto 1/m_1$.
(Note that the decay rate is related to the inverse of the 
corresponding mass gap.)
On the contrary,
the numerical diagonalization method
is suitable for investigating the spectrum
\cite{Lee01}.
However,
in practice, the tractable system size with the diagonalization method
is severely restricted;
the numerical diagonalization
requires a huge computer-memory space.
To cope with this difficulty,
in Ref. \cite{Lee01},
the authors truncated the number of Hilbert-space bases
(stochastic diagonalization method)
to calculate
the spectrum approximately.
The result appears to be consistent with the
above-mentioned Monte Carlo result.
(However, as indicated in Table 1 of
Ref. \cite{Lee01},
the results
$m_2/m_1$ scatter, depending on the distance from the critical point.
In this paper, we carry out systematic
finite-size-scaling analysis to resolve 
such uncertainty.)

In this paper, we employ the numerical diagonalization 
(transfer matrix) method
to calculate $m_2/m_1$.
In order to cope with the difficulties mentioned above,
we make the following modifications.
First, we improve the finite-size-scaling behavior.
That is, we reduce corrections to finite-size scaling
by tuning plaquette-type
(next-nearest-neighbor and four-spin)
interactions.
Such a reduction of corrections for the $d=3$ 
Ising model was pursued in Refs. \cite{Chen82,Blote96,Nishiyama06},
where the criticality of the $d=3$ Ising model was investigated.
(Similar attempt was made in a lattice-field-theoretical
context \cite{Symanzik83a,Symanzik83b,Ballesteros98,Hasenbusch99,Hasenbusch00}.)
To be specific, we follow the formalism advocated in Ref. \cite{Nishiyama06},
where the Hamiltonian reads
\begin{equation}
\label{Hamiltonian}
H=
-J_{NN} \sum_{\langle i,j\rangle}S_i S_j
-J_{NNN} \sum_{\langle\langle i,j\rangle\rangle}S_i S_j
-J_\Box \sum_{[i,j,k,l]}S_i S_j S_k S_l
-H \sum_i S_i
.
\end{equation}
Here, the Ising spins $\{ S_i=\pm 1 \}$ are placed at the
cubic-lattice points $i$,
and the summations,
$\sum_{\langle i,j \rangle}$,
$\sum_{\langle\langle i,j \rangle\rangle}$,
and $\sum_{[ i,j,k,l ]}$,
run over all nearest-neighbor pairs, next-nearest-neighbor spins, and
round-a-plaquette spins, respectively.
The parameters $\{ J_\alpha \}$ are the corresponding coupling constants.
As usual, $H$ denotes the uniform magnetic field.
[We set the temperature as a unit of energy ($T=1$) throughout this paper.]
We survey the one-parameter subspace
\begin{equation}
\label{parameter_space}
(J_{NN},J_{NNN},J_\square)=J_{NN}
 \left(
              1,\frac{\tilde{J}_{NNN}}{\tilde{J}_{NN}},
                \frac{\tilde{J}_\square}{\tilde{J}_{NN}}
   \right)
,
\end{equation}
containing a 
renormalization-group fixed point,
\begin{equation}
\label{fixed_point}
(\tilde{J}_{NN},\tilde{J}_{NNN},\tilde{J}_\square)=
(0.108982866643  5 
,
0.04 45777727956 
,
-0.0065117950492 
)  ,
\end{equation}
at $J_{NN}=\tilde{J}_{NN}$;
the fixed point was determined approximately
for a real-space-decimation transformation
\cite{Nishiyama06}.
Around the fixed point,
the finite-size corrections may cancel out
because of the absence of irrelevant interactions
\cite{Blote96,Nishiyama06}.
Second, we carry out two types of finite-size-scaling analyses
for $H\ne0$ and $H=0$.
Such independent information allows us to estimate $m_2/m_1$
systematically.
Last, we utilize Novotny's transfer-matrix method
\cite{Novotny90,Nishiyama04},
which enables us to treat an arbitrary (integral) number of 
spins $N=5,6,\dots,15$
constituting a unit of the transfer-matrix slice;
note that conventionally,
the number of spins is
restricted within $N=4,9,16,\dots$.

The ratio $m_2/m_1$ was considered analytically in Ref. \cite{Caselle02}.
In the following,
we outline the argument in order to elucidate the
underlying physics.
The authors calculate
the four-point vertex (Bethe-Salpeter equation) 
of the $\phi^4$ model perturbatively
up to the second order;
the analyticity of vertex contains information on the bound state.
As a result, they expressed $m_2/m_1$
in terms of 
the coupling constant at the critical point;
the critical coupling itself is
a nontrivial parameter.
By using the critical coupling
\cite{Caselle96} determined numerically,
the authors arrive at a convincing result
$m_2/m_1=1.828(3)$ at the leading order.
However,
at the next-leading order,
the ratio $m_2/m_1$ turned out to be negative,
indicating that $m_2/m_1$ is highly non-perturbative.
Nevertheless, their consideration 
reveals
a close relationship between
$m_2/m_1$ and
the critical coupling (critical amplitude).

In fairness,
it has to be mentioned that the spectral structure 
of the $d=3$ Ising model is considered in Refs.
\cite{Weigel99,Weigel00,Deng02}.
Here, the relationship between
the scaling dimension and the corresponding excitation gap
is investigated for various geometries and boundary conditions
via the
conformal invariance.
In the present paper, we dwell on the
off-critical excitations 
having a physical interpretation as magnons.

The rest of this paper is organized as follows.
In Sec. \ref{section2},
we present the simulation results for $m_2/m_1$.
We also explain the simulation scheme.
In the last section, we present the summary and discussions.

\section{\label{section2} Numerical results}

In this section, we present the numerical results for 
the $d=3$ Ising model, Eq. (\ref{Hamiltonian}),
with the extended interactions, Eq. (\ref{parameter_space}).
We employ Novotny's transfer-matrix method 
\cite{Novotny90,Nishiyama04} for finite clusters
with $N=5,6,\dots,15$ spins.
To begin with, we outline the simulation algorithm.

\subsection{\label{2_1}Simulation method: Novotny's technique}

In this section, 
we explain the simulation scheme \cite{Novotny90,Nishiyama04,Nishiyama06} briefly;
we refer readers to consult with
Ref. \cite{Nishiyama04} for details.

\begin{figure}
\includegraphics[width=100mm]{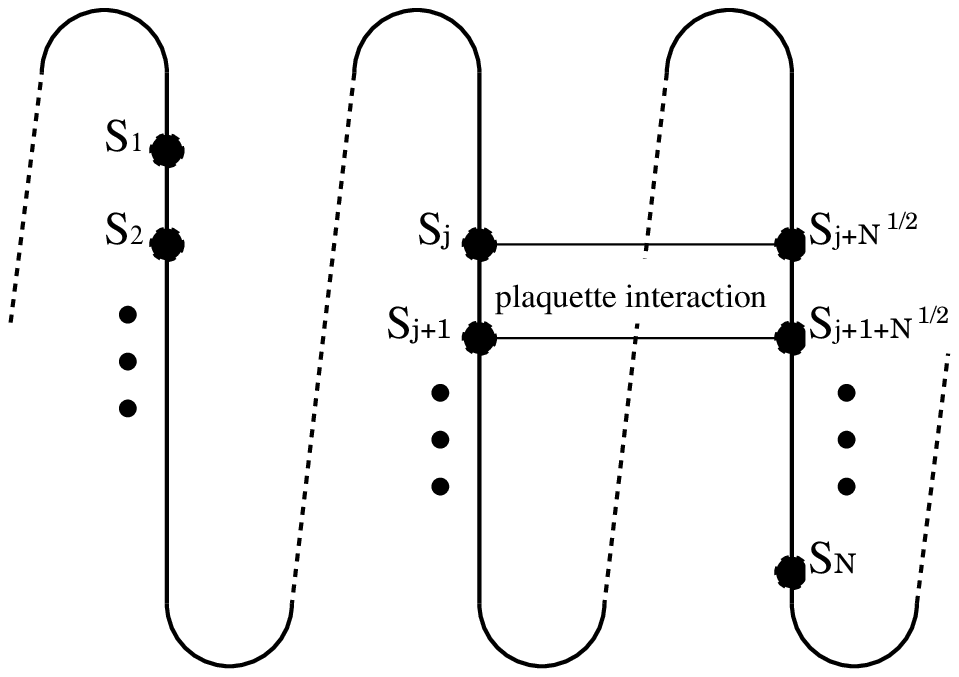}%
\caption{
\label{figure1}
A schematic drawing of 
a transfer-matrix unit of the
$d=3$ Ising model (\ref{Hamiltonian}).
As indicated, the spins constitute a 
$d=1$-dimensional (zigzag) alignment $\{ S_i \}$ 
($i=1,2, \dots ,N$),
and the dimensionality is lifted to $d=2$ by the bridges over the 
$(N^{1/2})$th-neighbor pairs.
(A transfer-matrix unit for $d=3$ should have a $d=2$ structure.)
Full details of the algorithm are presented in Ref. \cite{Nishiyama04}.
}
\end{figure}

A schematic drawing of a unit of the transfer-matrix slice
(cross-section of the transfer-matrix bar)
is presented in Fig. \ref{figure1}.
Basically,
the spins constitute a $d=1$-dimensional 
(zigzag) alignment $\{ S_i \}$ 
($i=1,2,\dots,N$).
The dimensionality is lifted to $d=2$
by the bridges over the $(N^{1/2})$th neighbor pairs.
Because the basic structure is one-dimensional,
we are able to construct the transfer-matrix unit
with an arbitrary (integral) number of spins;
note that conventionally, the number of spins for the $d=2$ structure is
restricted within $N=4,9,16,\dots$ (quadratic numbers).
The linear dimension $L$ is given by
\begin{equation}
L=N^{1/2}   ,
\end{equation}
because the $N$ spins form a rectangular cluster.
We treated the system sizes
$N=5,6,\dots, 15$ with this technique.

Our main concern is to calculate the excitation gap.
We perform the numerical diagonalization within the
subspace $k=0$ ($k$: the wave number).
Additionally,
we consider the spin-inversion-symmetry index $\pm$
in the case of $H=0$.
First,
for $H \ne 0$,
the $i$th excitation gap is given by the formula
\begin{equation}
m_i = \ln (\lambda_0/\lambda_i)  .
\end{equation}
Here, the parameters $\{ \lambda_i \}$
are the eigenvalues of the transfer matrix,
satisfying the inequality, $\lambda_0 > \lambda_1 > \lambda_2 > \dots$.
Second,
in the case of $H=0$,
the first and second excitation gaps are given by
\begin{equation}
\label{H0_gap1}
m_1  = \ln (\lambda^+_0/\lambda^-_1)    
\end{equation}
and
\begin{equation}
\label{H0_gap2}
m_2  = \ln (\lambda^+_0/\lambda^+_1)   ,
\end{equation}
respectively,
with the transfer-matrix eigenvalues $\{ \lambda^\pm_i \}$ 
specified by the spin-inversion-symmetry index
$\pm$.

In practice,
we need to modify Novotny's method 
in order to incorporate the
extended interactions such as the next-nearest-neighbor
and four-spin couplings
\cite{Novotny90}.
We refer readers to consult with Ref. \cite{Nishiyama04} for  
a detailed description of the algorithm.

Lastly, to avoid confusion,
we address a remark on the fixed point, Eq. (\ref{fixed_point}).
The fixed point is not used to analyze the criticality such as
the mass-gap ratio $m_2/m_1$.
We utilize the conventional finite-size-scaling method.
We perform the computer simulation around the fixed point,
where corrections to scaling cancel out satisfactorily;
see the Introduction.
In this sense, the fixed-point analysis is a preliminary
to the finite-size scaling.
Owing to the suppression of finite-size errors,
we are able to analyze the simulation data systematically.

\subsection{\label{section2_2}
Finite-size-scaling analysis of $m_2/m_1$: $H\ne 0$}

In this section, we analyze the simulation data 
under an infinitesimal magnetic field $H \ne 0$.
The magnetic field stabilizes the bound state.
(The case of $H=0$ is considered in the next section.)

To begin with, we consider the finite-size scaling
for the mass-gap ratio
\begin{equation}
\label{scaling_relation}
\frac{m_2}{m_1}=f \left( (J_{NN}-J^*_{NN})L^{1/\nu},HL^{y_h} \right)  ,
\end{equation}
with the critical point 
$J^*_{NN}=0.11059  $ \cite{Nishiyama06}.  
The critical indices, $\nu$ and $y_h$,
are taken from the Monte Carlo estimates
$\nu=0.63020$ and $y_h=2.4816$ 
(3d-Ising universality) reported in Ref. \cite{Deng03}.
The above scaling relation, Eq. (\ref{scaling_relation}),
stems from the fact that the mass-gap ratio 
is a dimensionless (scale invariant) quantity.

Based on the above scaling relation (\ref{scaling_relation}),
in Fig. \ref{figure2},
we presented a plot,
$(J_{NN}-J^*_{NN})L^{1/\nu}$-$m_2/m_1$,
for
various $J_{NN}$, 
 $N=5,6,\dots,15$,
and
the fixed scaling parameter $HL^{y_h}=4$;
afterward, we argue the constraint,
$HL^{y_h}=4$, in detail.
We see that the data collapse into a scaling curve.
We stress that
there are no fitting parameters in the finite-size scaling of Fig. \ref{figure2}.
Actually, all scaling
parameters, $J^*_{NN}$, $\nu$, and $y_h$, are taken from the existing values,
as mentioned above.

Noticeably enough, there appears
a plateau with $m_2/m_1 \approx 1.8$
extending in the low-temperature side, $J_{NN}-J^*_{NN}>0$.
This feature indicates
an existence of a bound state with $m_2/m_1\approx 1.8$
in the broken-symmetry phase.
As mentioned in the Introduction,
a properly scaled magnetic field is useful to
observe
the bound state clearly.

The plateau of Fig. \ref{figure2} yields the mass-gap ratio.
We read off $m_2/m_1$ at $(J_{NN}-J^*_{NN})L^{1/\nu}=0.3$ for each system size,
and plotted the values against $1/L^2$ in Fig. \ref{figure3}.
(Afterward, we consider the validity of the abscissa scale $1/L^2$.)
The least-squares fit to the data for $N \ge 7$ yields an
estimate $m_2/m_1=1.835(4)$ in the thermodynamic limit.

Lastly, we address a few remarks.
First,
we consider the scaled magnetic field $H=4/L^{y_h}$.
Owing to the scaling,
the second argument of the scaling relation, Eq. (\ref{scaling_relation}),
is kept invariant;
namely, the presence of magnetic field becomes implicit under the scaling. 
On one hand, the magnetic field
plays a significant role to stabilize the bound state.
Second, we explain why the coefficient $A=4$ in the scaling $H=A/L^{y_h}$
is an optimal one.
We surveyed various values of $A$,
and found that for $A>5$, the plateau inclines;
namely, it becomes ambiguous where we measure the plateau hight.
For $A<3$, eventually, there occur
successive level crossings, smearing out the plateau structure;
the magnetic field $H$ may be too weak to stabilize the plateau (bound state).
Last, we consider the extrapolation scheme $1/L^2$
(abscissa scale in Fig. \ref{figure3}).
As shown in the figure,
the data for $H \ne 0$,
and $H=0$ 
are consistent with each other.
Moreover, using the $\nu$ data reported in Ref. \cite{Nishiyama06},
we arrive at a convincing result $\nu=0.6314(14)$
through the $1/L^2$ extrapolation scheme; 
the result is comparable with that of
Monte Carlo $\nu=0.63020(12)$ \cite{Deng03}.
(On the contrary, 
in Ref. \cite{Nishiyama06}, a rather naive extrapolation scheme $1/L$
was used, resulting in $\nu=0.6245(28)$.)
In principle, the finite-size corrections should obey the 
formula
$1/L^\omega$ with $\omega=0.821(5)$ \cite{Deng03}.
That is, the convergence of our result
is rather accelerated.
This fact indicates that
the finely tuned coupling constants, Eq. (\ref{parameter_space}),
suppress
the finite-size corrections significantly.
So far, the technique has been applied
to investigating
the equilibrium properties
\cite{Blote96,Nishiyama06}.
Our result indicates that
the technique is applicable to 
the spectral property as well.

%

\subsection{\label{section2_3}
Finite-size-scaling analysis of $m_2/m_1$: $H=0$}

In this section, we consider the case of $H=0$.
Here, we have to make the following replacement
\begin{equation}
m_3/m_2 \to m_2/m_1       ,
\end{equation}
because the ground state is doubly degenerated
owing to the spontaneous symmetry breaking;
namely, the first-excitation gap $m_1=0$ does not make any sense.
(More specifically,
the excitations are given by Eqs. (\ref{H0_gap1}) and (\ref{H0_gap2}).)
Actually, to avoid this complication, we applied a magnetic
field, and resolve the degeneracy in the above section.

In Fig. \ref{figure4},
we present the scaling plot,
$(J_{NN}-J^*_{NN})L^{1/\nu}$-$m_2/m_1$,
for various $J_{NN}$, $N=5,6,\dots,15$ and the fixed 
magnetic field $H=0$.
The scaling parameters are the same as those of Fig. \ref{figure2}.
The data for $N \ge 7$ collapse into a scaling curve satisfactorily.

From Fig. \ref{figure4}, we read off $m_2/m_1$ at
$(J_{NN}-J^*_{NN})L^{1/\nu}=0$ (critical point)
for each $N$, and 
plotted them against $1/L^2$
in Fig. \ref{figure3}.
The least-squares fit to the data for $N \ge 7$
yields an estimate $m_2/m_1=1.841(4)$ in the thermodynamic limit.
The result is consistent with that of $H\ne0$ (Sec. \ref{section2_2}).
In the last section,
we address a concluding remark.

\subsection{Scaling analysis of the ordinary Ising model $J_{NNN}=J_\Box=0$}

In this section, tentatively,
we turn off the extended interactions,
$J_{NNN}=J_\Box=0$.
That is, we consider the ordinary Ising model, which is relevant to
the preceding Monte Carlo study \cite{Caselle99}.

In Fig. \ref{figure5}, we present the plot,
$(J_{NN}-J^*_{NN})L^{1/\nu}$-$m_2/m_1$,
with $J^*_{NN}=0.22165455$ \cite{Deng03},
$N=5,6,\dots,15$,
$J_{NN}=J_\Box=0$, and $H=0$; 
the other scaling parameters are the same as those of Fig. \ref{figure4}.

We see that the data scatter as compared to those of 
Fig. \ref{figure4}, where we incorporated the extended interactions.
This result demonstrates that the extended interactions suppress the
finite-size errors significantly.
Such a suppression of corrections was observed in Refs. \cite{Blote96,Nishiyama06},
where the thermal equilibrium properties, rather than
the spectral ones, were explored with this technique.

As mentioned above,
the Monte Carlo simulation \cite{Caselle99}
was performed at $H=0$ and $J_{NN}=J_\Box=0$.
We argue an implication of the present simulation data
as to the preceding Monte Carlo study.
Our scaling analysis, Fig. \ref{figure5},
 indicates that the
scaling regime,
where we observe correct mass-gap ratio $m_2/m_1\sim 1.8$, is extremely
narrow $(J_{NN}-J^*_{NN})L^{1/\nu}<0.2$.
On the other hand, the Monte Carlo simulation
was performed at $(J_{NN}-J^*_{NN})L^{1/\nu} \approx 2.16$ and $2.46$ 
with $(J_{NN},L)=(0.23142,30)$ and $(0.2275,45)$, respectively;
those parameters are
rather out of the scaling regime.
Nevertheless, 
in such close vicinity
to the critical point, 
the notorious critical-slowing-down problem would emerge.
Possibly,
the data under $H \ne 0$ 
may provide valuable information
on $m_2/m_1$ 
as demonstrated in 
Sec. \ref{section2_2}.

\section{\label{section3}Summary and discussions}

The bound-state energy $m_2$ of the $d=3$ Ising model
in the broken-symmetry phase,
$J_{NN}-J^*_{NN}>0$,
is investigated numerically.
The mass-gap ratio $m_2/m_1$
($m_1$: elementary excitation gap) 
is a universal constant
in the vicinity of the critical point.
We employed 
Novotny's transfer-matrix method \cite{Novotny90,Nishiyama04}
to treat a variety of system sizes
$N=5,6,\dots,15$.

In order to improve the finite-size-scaling behavior,
we extended the interactions, Eq. (\ref{Hamiltonian}),
following Ref. \cite{Nishiyama06}.
Owing to this improvement,
we attain improved finite-size corrections 
(Fig. \ref{figure4}),
as compared to those of the ordinary Ising model 
(Fig. \ref{figure5});

\begin{figure}
\includegraphics[width=100mm]{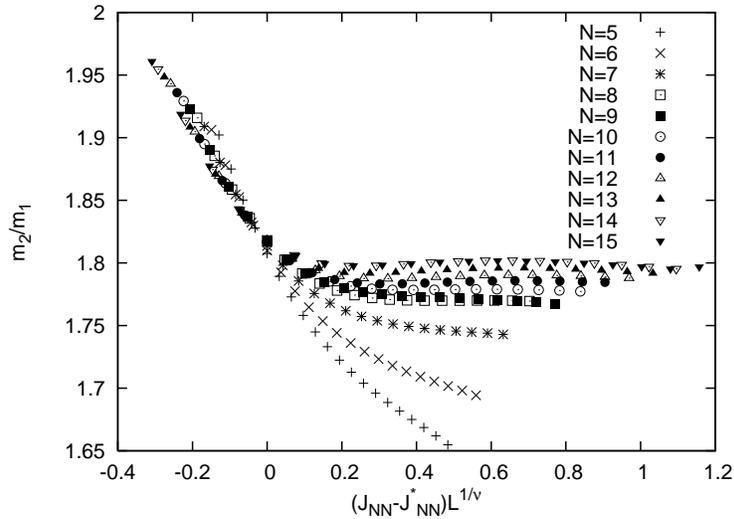}%
\caption{
\label{figure2}
Finite-size-scaling plot,
$(J_{NN}-J^*_{NN})L^{1/ \nu}$-$m_2/m_1$, for 
the $d=3$ Ising model
under the scaled magnetic field 
$H=4/L^{y_h}$ is shown.
Here, we set
$J^*_{NN}=0.11059$ \cite{Nishiyama06},
and $(\nu,y_h)=(0.63020,2.4816)$ (3d-Ising universality) 
\cite{Deng03}.
From the plateau in the low-temperature side,
we estimate the mass-gap ratio $m_2/m_1$ 
for each $N$; see text for detail.
In Fig. \ref{figure3}, the mass-gap ratio is plotted
against $1/L^2$.
}
\end{figure}

\begin{figure}
\includegraphics[width=100mm]{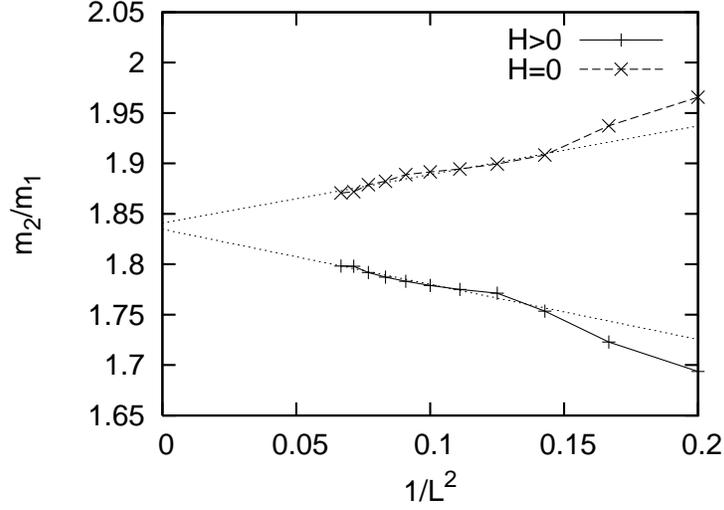}%
\caption{
\label{figure3}
The mass-gap ratio $m_2/m_1$ is plotted for $1/L^2$.
The symbols,
$+$ and $\times$, denote the data for $H \ne 0$
(Fig. \ref{figure2})
and $H=0$ 
(Fig. \ref{figure4}), respectively; see text for details.
The least-squares fit to these data
for the system sizes $N \ge 7$ yields 
$m_2/m_1=1.835(4)$ and $1.841(4)$, respectively, in the thermodynamic limit.
}
\end{figure}

\begin{figure}
\includegraphics[width=100mm]{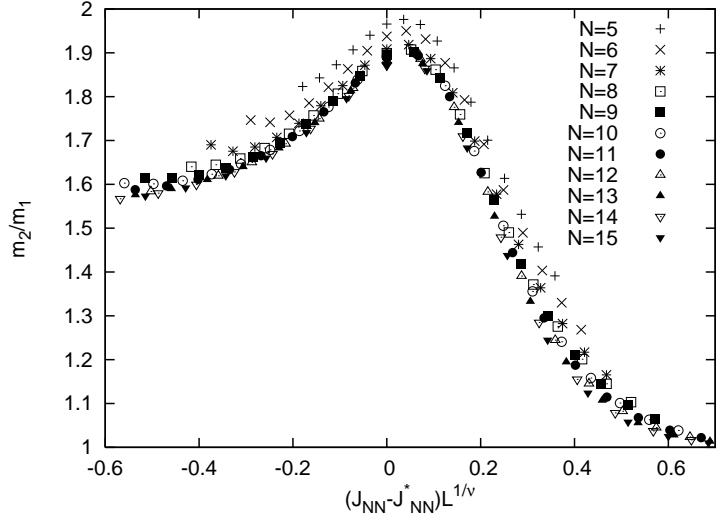}%
\caption{
\label{figure4}
Finite-size-scaling plot,
$(J_{NN}-J^*_{NN})L^{1/ \nu}$-$m_2/m_1$, for 
the $d=3$ Ising model (\ref{Hamiltonian})
without magnetic field $H=0$
is shown.
The scaling parameters are the same as those of Fig. \ref{figure2}.
At $J_{NN}-J^*_{NN}=0$, we obtain the mass-gap ratio $m_2/m_1$
for each system size $N$;
see Fig. \ref{figure3}.
}
\end{figure}

\begin{figure}
\includegraphics[width=100mm]{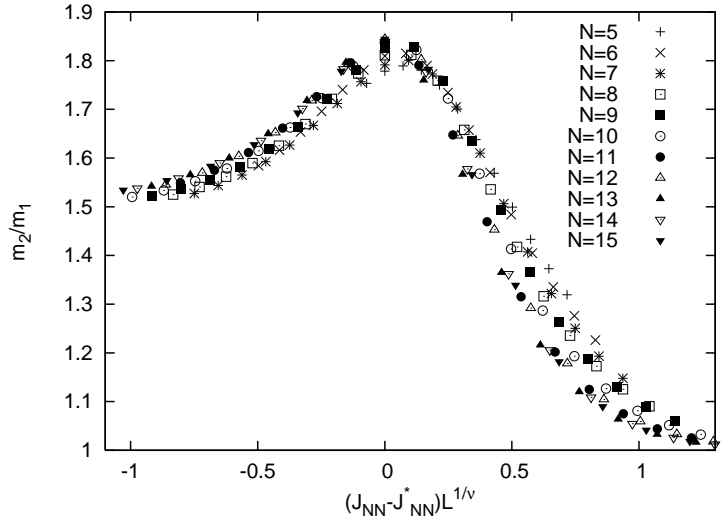}%
\caption{
\label{figure5}
Finite-size-scaling plot,
$(J_{NN}-J^*_{NN})L^{1/ \nu}$-$m_2/m_1$, for 
the $d=3$ Ising model (\ref{Hamiltonian})
without magnetic field $H=0$
is shown.
Here, we tentatively turned off the extended interactions $J_{NNN}=J_\Box=0$.
The critical point is set to
$J^*_{NN}=0.22165455$ \cite{Deng03}, 
and the other scaling parameters are the same as those of Fig. \ref{figure2}.
We see that the data scatter 
as compared to those of Fig. \ref{figure4},
where we incorporated the extended interactions.
}
\end{figure}

In Figs. \ref{figure2} and \ref{figure4},
we made different approaches to the critical point
through $H\ne0$ and $H=0$, respectively.
As shown in Fig. \ref{figure3},
these results provide lower and upper bounds for $m_2/m_1$,
respectively.
Combining these results,
we estimate the mass-gap ratio 
\begin{equation}
m_2/m_1=1.84(3)
   .
\end{equation}
The result is consistent with that of Monte Carlo 
$m_2/m_1=1.83(3)$ \cite{Caselle99}.
However, as argued in Sec. \ref{section2_3},
there arises a subtle discrepancy as to the scaling regime.

So far, the technique to suppress finite-size corrections
has been utilized to survey the 
thermal-equilibrium (ground state) properties
\cite{Blote96,Nishiyama06}.
The present analysis demonstrated that this
technique is also applicable to the excitation spectrum.
Likewise, the technique would be of use to investigate
the dynamical properties such as the
spectral intensity and the
relaxation to thermal equilibrium.
This problem is addressed in future study.

\begin{acknowledgments}
This work was supported by a Grant-in-Aid 
(No. 18740234) from Monbu-Kagakusho, Japan.
\end{acknowledgments}


\end{document}